\documentclass[epj]{svjour}
\usepackage{graphics}
\usepackage{epsfig}
\usepackage{amssymb}
\usepackage{amsmath}
\usepackage[pdftex,a4paper,colorlinks=true,citecolor=blue,urlcolor=blue]{hyperref}
\usepackage[numbers,sort&compress]{natbib}
\begin{document}
\title{A new study of the $^{10}$B(p,$\alpha_1 \gamma$)$^{7}$Be reaction from 0.35 to 1.8 MeV}
\author{A.\,Caciolli \inst{1,2} \thanks{email: caciolli@pd.infn.it} \and 
	    R.\,Depalo \inst{2} \and
	    V.\,Rigato \inst{3}
}                     % Do not remove
\offprints{A. Caciolli}          % Insert a name or remove this line
\authorrunning{A.\,Caciolli \textit{et al.}}
\institute{Universit\`a degli Studi di Padova, Dipartimento di Fisica e Astronomia, via Marzolo 8, 35131 Padova, Italy
\and
INFN Sezione di Padova, via Marzolo 8, 35131 Padova, Italy
\and
INFN Laboratori Nazionali di Legnaro, via dell'Universit\`a 2, 35020 Legnaro, Padova, Italy}
\date{Draft \today}
\abstract{
The quantification of isotopes content in materials is extremely important in many  research and industrial fields. Accurate determination of boron concentration is very critical in semiconductor, superconductor and steel industry, in environmental and medical applications as well as in nuclear and astrophysics research. The detection of B isotopes and of their ratio in synthetic and natural materials may be accomplished by gamma spectroscopy using the $^{10}$B(p,$\alpha_1 \gamma$)$^7$Be and $^{11}$B(p,$\gamma$)$^{12}$C reactions at low proton energy. 
Here, the $^{10}$B(p,$\alpha_1 \gamma$)$^7$Be cross section is reported in the center of mass energy range  0.35 to 1.8~MeV. The  $E_\gamma$=~429~keV $\gamma$ rays were detected at 45$^\circ$ and 90$^\circ$ using a NaI(Tl) and an HPGe detectors, respectively. In the presented energy range, previous cross sections data revealed discrepancies and normalisation issues. Existing data are compared to the new absolute measurement and discussed. The present data have been subtracted from a previous measurement of the total cross section to derive the contribution of the  $\alpha_0$ channel.
%That method is blind to the possible branching cascade and only this subtraction will separate the two contribution.
%\PACS{
   %  {25.60.Dz}{Interaction and reaction cross sections}   \and
   % {29.20.Ba}{Electrostatic accelerators} \and
   %  {29.30.Kv}{X- and $\gamma$-ray spectroscopy}
   % } % end of PACS codes%
} %end of abstract
\maketitle
%

%=============================================================================
\section{Introduction}
\label{sec:intro}
%=============================================================================

The use of accelerated MeV ion beams for characterising the elemental content and depth profiles in materials and interfaces started in the early sixties  \cite{Mayer-book-1977,Nastasi-book-1995} and is a powerful analytical method in many  research and technological fields known as Ion Beam Analysis (IBA). More and more robust models and codes allowed the analysis of the elastic and inelastic processes in surface analysis \cite{RUMP,SIMNRA}. 
 IBA techniques have become more powerful including self-consistent analysis of energetic photon emission spectra together with particle scattering (RBS, EBS, ERD, NRA) spectra \cite{Jeynes12-NIMB}. 
A review of the analytical techniques may be found in \cite{Jeynes11-RAST}.
Nowadays, these techniques are well established and their development is related to the increase of data collection automatisation and to more and more accurate measurement of the cross section involved in the analyses. As a matter of fact, still many reaction cross sections are provided in literature and in data-bases \cite{IBANDL} with high error bars or with discrepancies between the different studies. 
Combining robust computational codes with the accurate cross sections data, the content, depth profile, and ultimately the detection limit of a given element or isotope in a specific matrix may be identified quite precisely. 
Boron is an important contaminant and dopant in semiconductor industry and is of fundamental importance in many other research, medical, environmental, and industrial fields, therefore studies on its quantification are still actual. A review of analytical and sample preparation methods for the determination of B in materials has been published in 1997 \cite{Sah97-MJ}. Traditionally, boron analysis with MeV ion beams is accomplished using a combination of (p,p) elastic backscattering and Nuclear Reaction methods with high positive Q values, such as (d,p), (d,$\alpha$), (p,$\alpha$), ($^3$He,p), ($^3$He,$\alpha$) reactions \cite{Nastasi-book-1995}. 
More recently Particle Induced Gamma-Ray Emission (PIGE) has been used \cite{Mateus07-NIMB,Moro14-AIPCP} for B analysis using the $^{10}$B(p,$\alpha_1 \gamma$)$^7$Be reaction. 

Below 4.5 MeV, $^{10}$B(p,$\alpha$)$^7$Be can proceed emitting $\alpha$ particles directly to the ground state (the so called $\alpha_0$ channel) or to the 429 keV excited state of $^7$Be ($\alpha_1 \gamma$ channel). In the $\alpha_1 \gamma$ channel a $\gamma$ ray with $E_\gamma$~=~429 keV is also emitted. The $\alpha_0$ channel has been studied deeply at energies below 2 MeV down to 5 keV (see \cite{Spitaleri17-PRC} and references therein). The present work is focused on the $\alpha_1 \gamma$ channel. The existing data of the $^{10}$B(p,$\alpha_1 \gamma$)$^7$Be reaction  are compared and discussed in section \ref{sec:state_art}. The experimental setup and data analysis are presented in section \ref{sec:setup}. The cross section results are presented in section \ref{sec:discussion}. The present data have been used also to disentangle the $\alpha_0$ and $\alpha_1 \gamma$ channels in a recent activation measurement of the  $^{10}$B(p,$\alpha$)$^7$Be total cross section \cite{Caciolli16-EPJA}.

%=============================================================================
\section{State of the Art}
\label{sec:state_art}
%=============================================================================

A pioneering  measurement of the $^{10}$B(p,$\alpha_1 \gamma$)$^{7}$Be reaction was done by Brown et al. in 1951 \cite{Brown51-PR}. Their  purpose was to study the mirror nuclei  $^7$Li and $^7$Be.
In this experiment, the $\alpha$ particles emitted by the reaction were detected by a ionisation chamber and a scintillator after passing through an analysing magnet.
The $\gamma$ radiation was detected at 90$^\circ$ by using a Geiger-M\"uller detector. 
During the  experiment they  measured not only the cross sections of both the $\alpha_0$ and $\alpha_1 \gamma$ reaction channels for the $^{10}$B + p system, but also the elastic and inelastic scattering  on boron and lithium isotopes, spanning a proton energy range from 500 keV to 1.6 MeV.

In 1954, a subsequent paper from Day and Huus \cite{Day54-PR} reported new measurements for the $^{10}$B(p,$\alpha_1 \gamma$)$^{7}$Be,   $^{10}$B(p,p$'$ $\gamma$)$^{10}$B, and  $^{10}$B(p,$\gamma$)$^{11}$C with a NaI(Tl) 1.5"$\times$1.5" scintillator placed at 90$^\circ$ with respect to the beam direction.
They found a maximum of the $^{10}$B(p,$\alpha_1 \gamma$)$^{7}$Be cross section of 0.21$\pm$0.07 b at $E_p$ = 1.52 MeV, and compared that value with the one obtained by Brown \emph{et al.}  that was 0.14$\pm$0.03 b.
After this comparison, Day and Huus adopted the average of the two values ($\sigma$ = 0.16 b) in order to normalise their whole dataset, which is now reported in Fig. \ref{fig:cross_section} of this paper. 
%The authors then normalised the entire cross section curve to the average of the two above mentioned experimental values  ($\sigma$ = 0.16 b).
The data in \cite{Day54-PR} are the most used nowadays in ion beam analysis, despite the fact that the normalisation procedure was quite arbitrary.

Other experimental measurements were reported in the papers by Cronin \cite{Cronin56-PR} and Hunt et al. \cite{Hunt57-PR}. 
Cronin was  able to measure the angular distribution of out-coming particles in a range from 25$^\circ$ to 145$^\circ$ finding almost flat angular distribution for the $\alpha_1 \gamma$ channel at various energies around the resonance at 1.52 MeV (corresponding to 1.38 MeV in the center of mass).
The total cross section seems to be slightly lower than the Day and Huus data, but the wide energy steps do not allow a comparison at the resonance maximum.

On the contrary, the Hunt et al. dataset is a factor of two lower than that reported by Day and Huus and also in disagreement  with the Brown et al. data.

A recent work from Lagoyannis and coauthors \cite{Lagoyannis15-NIMB} investigated the reaction in an energy range from 2 to 5 MeV.
The data shown in this work slightly deviate from those reported in \cite{Day54-PR} in the overlapping region. Day and Huus claimed a problem at these energies due to the background produced by the 718 keV $\gamma$ ray from the inelastic proton scattering on $^{10}$B. This may explain the difference observed in the two datasets.
Lagoyannis et al., \cite{Lagoyannis15-NIMB} studied also the angular distribution of this reaction from 2 to 5 MeV claiming that the comparison between the eight different detection angles revealed no significant angular dependence of the cross section. 

It is clear that the picture of the data in literature is quite confusing (see Fig. \ref{fig:cross_section}) and, in particular, all datasets show discrepancies with each other. 
In addition, we should point out that the normalisation procedure adopted in \cite{Day54-PR} reduced this discrepancy, while their absolute results increase the spread in the data in literature.

The present paper provides new, higher precision absolute cross sections in the same energy range as \cite{Day54-PR}, improving the present understanding of the discrepancies. 

%=============================================================================
\section{Experimental Setup and Analysis}
\label{sec:setup}
%=============================================================================

\begin{figure}[!tbh]
\centering
\includegraphics[width=\columnwidth]{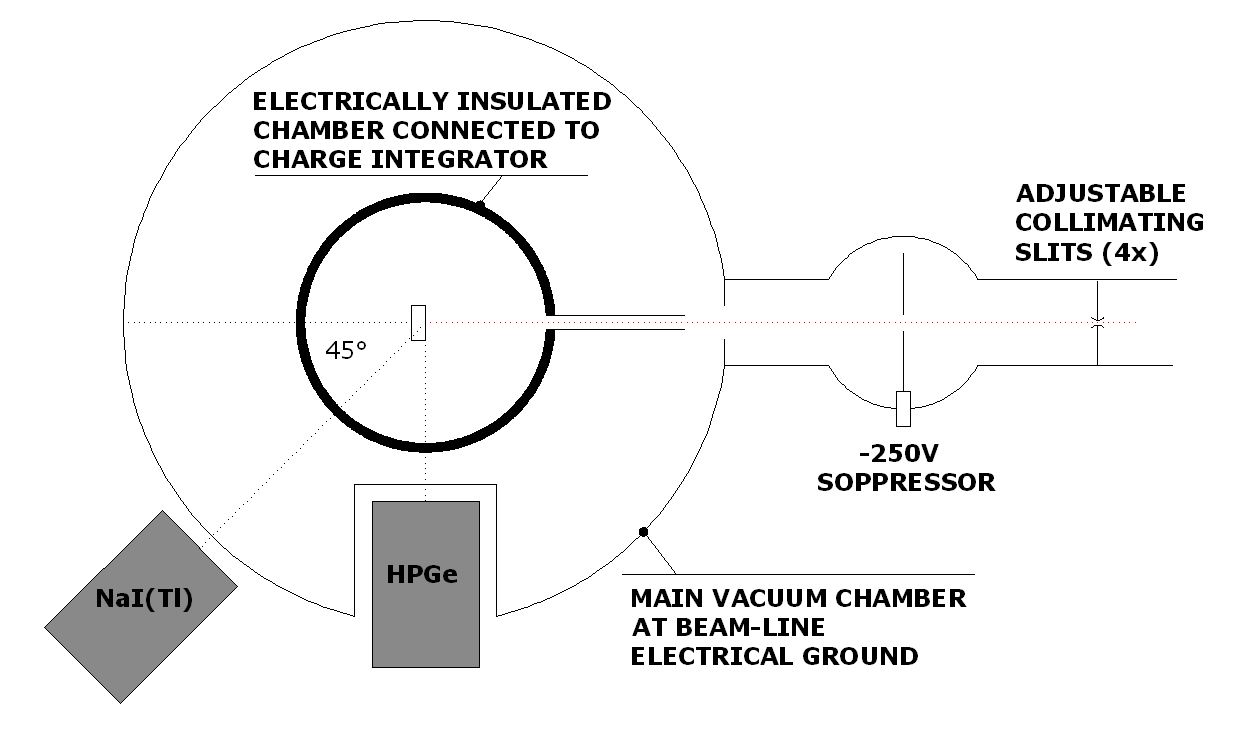}
\caption{A schematic view of the setup used during the present experiment (see text for details).}
\label{fig:setup_scheme}
\end{figure}

The experiment was carried out at the AN2000 Van de Graaff accelerator Laboratory of the Laboratori Nazionali di Legnaro of INFN. 
The setup was installed on the 60$^\circ$ beam-line. The 14 cm diameter and 12 cm height scattering chamber is installed inside a bigger chamber as depicted in Fig. \ref{fig:setup_scheme}. It is made of aluminum and it is electrically isolated from the other chamber and the beam-line. 
On the bottom of the scattering chamber a labyrinth path with adequate conductance allows the chamber to be pumped.
The beam enters into the chamber passing through a 8 mm diameter tube placed at 15 cm from the target. In this way, the charge losses are minimised. The beam current and integrated beam charge have been determined with an ORTEC 439 current integrator using the scattering chamber as a Faraday cup. 
The scattering chamber/Faraday cup is properly aligned with the ion beam, whose size is of about 1 mm, as defined by 4 individually adjustable  collimating slits. Suppression of secondary electrons from external collimating slits is accomplished by applying a negative bias of about -250~V with respect to the beam-line pipe.
 
The proton energy has been defined by measuring the magnetic field of the analysing magnet that bends the beam to the 60$^\circ$ beam-line, using a NMR gauss-meter. The beam energy was calibrated during a previous experiment \cite{Caciolli16-EPJA} with the 992 and 632 keV $^{27}$Al(p,$\gamma$)$^{28}$Si resonances and then confirmed with subsequent experiments. The beam energy was calibrated  with a precision of 1 keV. It has to be noted that this uncertainty propagates to the cross section calculation. As a matter of fact, the experimental yield, defined as the number of reactions divided by the number of incident protons,  is related to the cross section by the equation:
%%==========================================
\begin{equation}\label{eq1}
Y(E_p) = \int^{E_p}_{E_p-\Delta E}\frac{\sigma(E)}{\epsilon_{eff}(E)}\mbox{d}E,
\end{equation}
%%==========================================  
where $\Delta E$ is the energy loss in the target and $\epsilon_{eff}(E)$ is the effective stopping power \cite{Iliadis15-Book} which is the stopping power weighted on the number of active isotopes in the target. The integral interval is calculated on the basis of the entrance energy and the target composition. This 1~keV uncertainty on the beam energy translates to an uncertainty of 1\% or less on the cross section in the whole explored energy range.

Boron powder, 92\% enriched in $^{10}$B, was used to produce the targets by $e$-beam evaporation. The films were deposited onto thin self-supporting carbon foils (about 20~$\mu$g/cm$^2$). Despite the starting nominal powder was only made of boron, contamination of nitrogen and oxygen could bond with boron during the handling of the powder and in the evaporation process itself.

%+++++++++++++++++++++++++++++++++++++++++++++++++++++++++++++++++++++++++++++++++
%+++++++++++++++++++++++++++++++++++++++++++++++++++++++++++++++++++++++++++++++++

Since the target composition is a crucial parameter and usually  the biggest source of uncertainty in absolute cross section measurements \cite{Caciolli12-EPJA},  a careful target characterization has been performed with proton-EBS \cite{Jeynes12-NIMB} at three different energies.
More precisely, a proton beam of three different energies, around 2~MeV, was used to bombard the targets and backscattered particles were detected by using a silicon detector (100 $\mu$m and 50 mm$^2$ active area) placed at 160$^\circ$.

\begin{figure}[!tbh]
\centering
\includegraphics[width=\columnwidth]{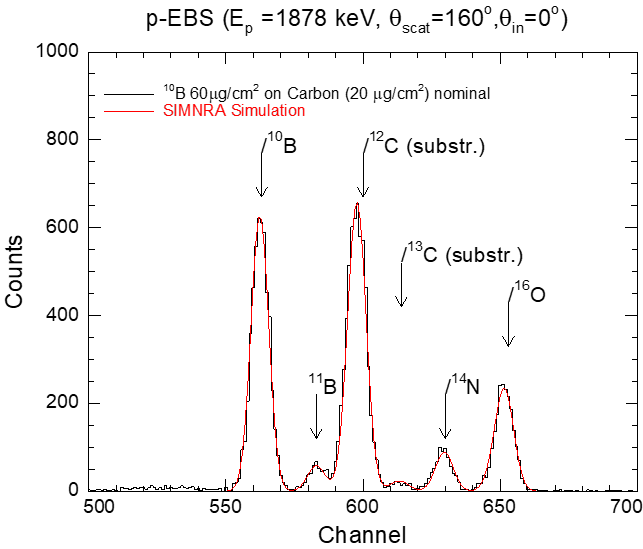}
\caption{A typical p-EBS spectrum analysed with SIMNRA 6.0 of one of the samples described in Table \ref{tab:target_analysis}. $\theta_{in}$ is the beam direction and $\theta_{scat}$ the angle of the silicon detector with respect to the beam direction.}
\label{fig:target_analysis}
\end{figure}

Three different nominal thicknesses were deposited for the experiment: 25, 60, and 100 $\mu$g/cm$^2$. 
For each nominal thickness 6 samples were produced. 
%This  number guarantees the homogeneity of the evaporation on all samples.
To characterise the targets and to determine possibile contamination inclusions, two samples for each bench were analysed finding  good agreement as shown in Table \ref{tab:target_analysis}. The EBS spectra were analysed with SIMNRA-6.0 \cite{SIMNRA} and RUMP \cite{RUMP} simulation softwares.
An example of the  elastic backscattered spectra is reported in Fig. \ref{fig:target_analysis}.
Oxygen and nitrogen contaminations were observed.
The averaged composition has been used in the analysis, while the maximum discrepancy has been adopted as a systematic uncertainty. 
%%%%%%%%%%%%%%%%%%%%%%%%%%%%%%%%%%%%%%%%%%%%%%%%%%%%%%%%%%%%%%%%%%%%%%%%%
\begin{table}[h]
\begin{center}
    \begin{tabular}{| l | c | c | c | c |}
    \hline
    Sample & Boron  & Oxygen  & Nitrogen & $^{10}$B  \\
    		& [10$^{15}$ $\frac{atoms}{cm^2}$]		&	[10$^{15}$ $\frac{atoms}{cm^2}$]	&	[10$^{15}$ $\frac{atoms}{cm^2}$]	&	[\%]	\\
\hline
\hline
     	25 & 296 & 217 & 36 & 92.9 \\ 
     	25 & 286 & 199 & 35 & 92.3 \\
	\hline
     	60 & 776 & 166 & 92 & 93.4 \\
     	60 & 780 & 167 & 93 & 93.5 \\
	\hline
     	100 & 1204 & 343 & 132 & 93.3 \\
     	100 & 1285 & 311 & 149 & 93.4 \\
\hline
        \end{tabular}
\caption{The results of the proton-EBS on two samples for each evaporation performed to obtain the three different thicknesses. Last column represents the $^{10}$B content with respect to  total boron in the samples.}\label{tab:target_analysis}
\end{center}
\end{table}
%%%%%%%%%%%%%%%%%%%%%%%%%%%%%%%%%%%%%%%%%%%%%%%%%%%%%%%%%%%%%%%%%%%%%%%%%
%+++++++++++++++++++++++++++++++++++++++++++++++++++++++++++++++++++++++++++++++++
%+++++++++++++++++++++++++++++++++++++++++++++++++++++++++++++++++++++++++++++++++
It has to be noted that this set of targets was used both for the present experiment and for the one in \cite{Caciolli16-EPJA}. In this experiment, only two targets were used with 60 and 100 $\mu$g/cm$^2$ nominal thicknesses, while for the one in \cite{Caciolli16-EPJA} all thicknesses were used. 

The reaction yield has been measured by using two $\gamma$-ray detectors placed at different angles: an High Purity Germanium (HPGe) detector of 50\% relative efficiency was installed at 90$^\circ$ with respect to the beam direction, while a 1 litre NaI(Tl) at 45$^{\circ}$. The signals of both detectors were recorded with a standard analog acquisition chain.

\begin{figure}[!tbh]
\centering
\includegraphics[width=\columnwidth]{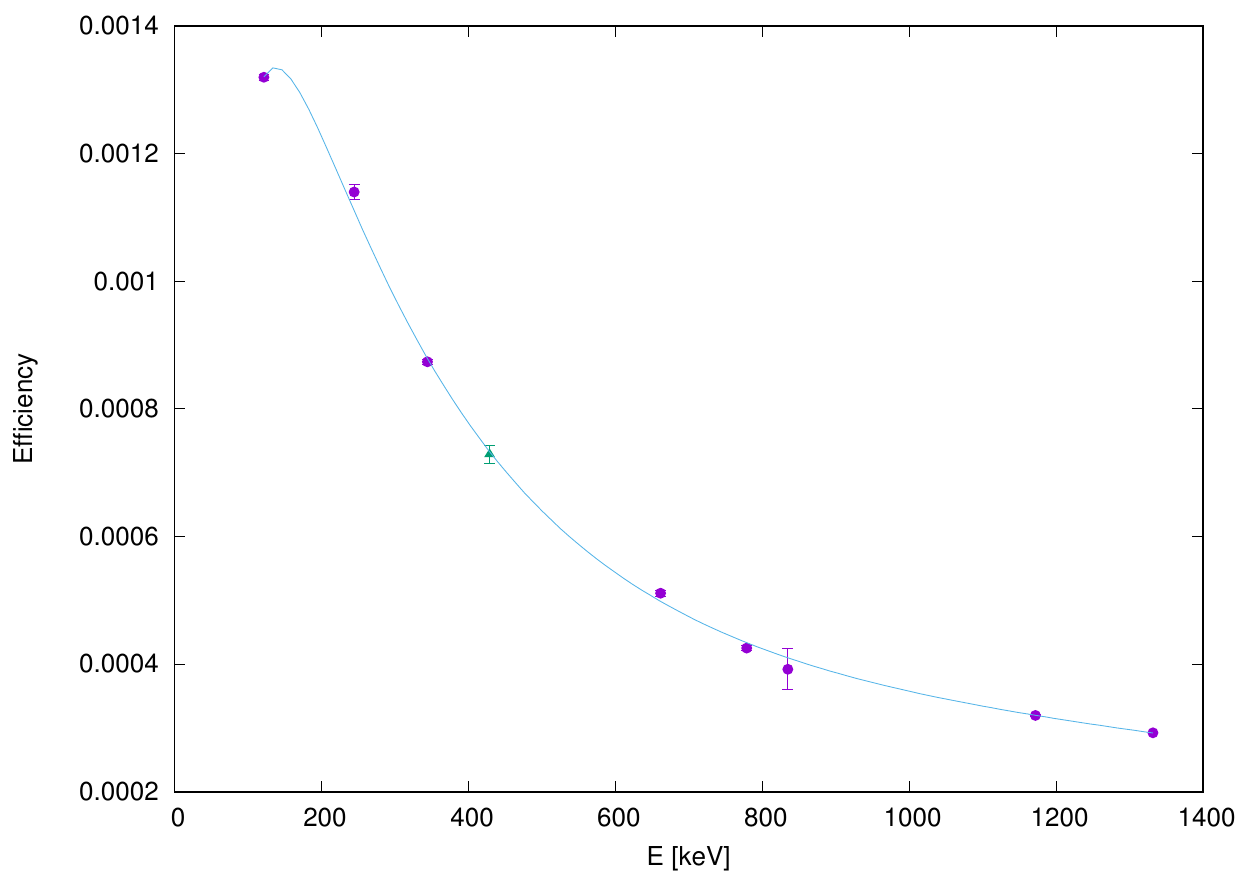}
\caption{The HPGe absolute efficiency curve used for the analysis. The points represent the experimental data while the line is the fit performed using the formula reported in \cite{Xhixha13-JRNC}. The triangle represents the efficiency used in the analysis with the uncertainty due to the fit.}
\label{fig:HPGe_efficiency}
\end{figure}

\begin{figure}[!tbh]
\centering
\includegraphics[width=\columnwidth]{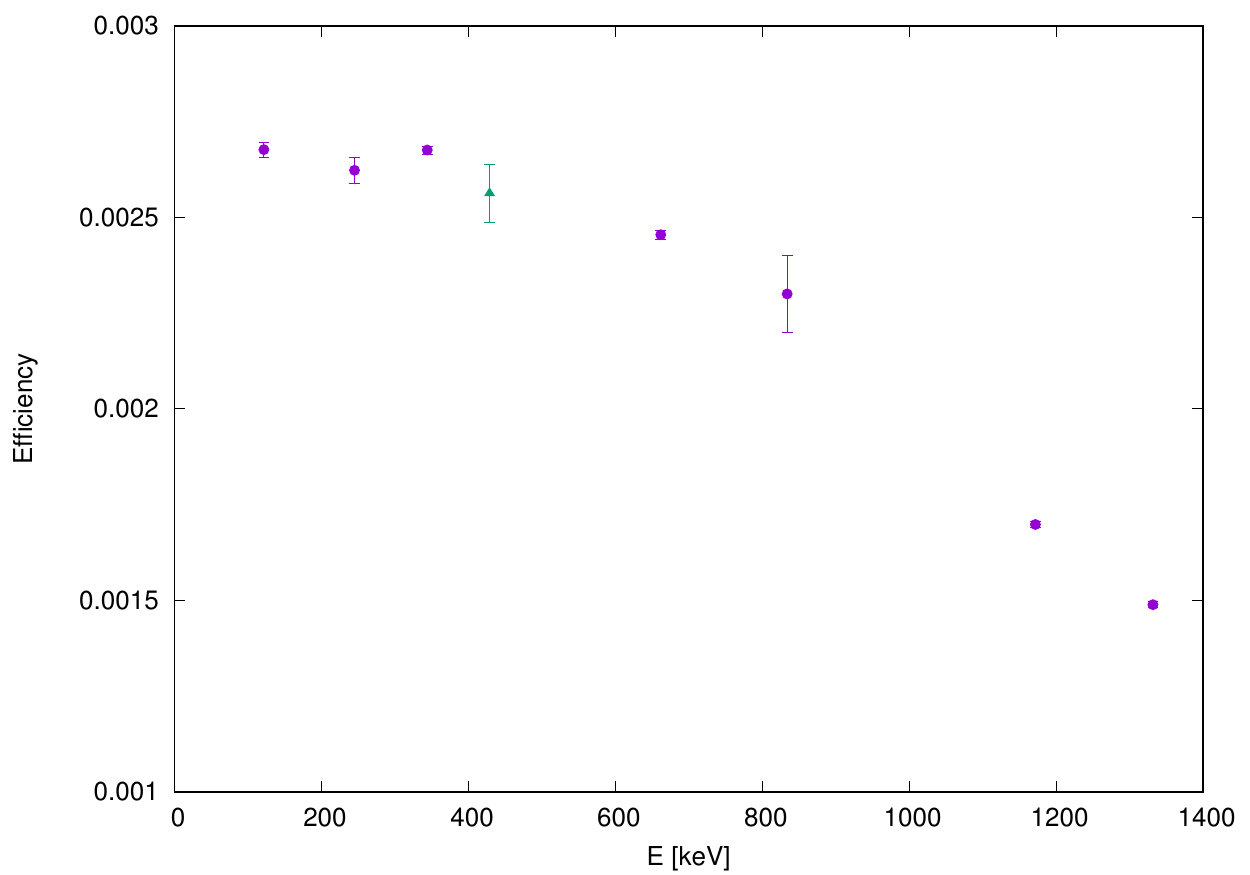}
\caption{The NaI absolute efficiency curve used for the analysis. The points represent the experimental data with their statistical uncertainties. The triangle represents the efficiency used in the analysis with the uncertainty due to the interpolation.}
\label{fig:NaI_efficiency}
\end{figure}

The two detectors have been calibrated in absolute efficiency by using the certified radioactive sources listed in Table \ref{tab:sources}.
%%%%%%%%%%%%%%%%%%%%%%%%%%%%%%%%%%%%%%%%%%%%%%%%%%%%%%%%%%%%%%%%%%%%%%%%%
\begin{table}[h]
\begin{center}
    \begin{tabular}{| l | c | c | c |}
    \hline
    Nuclide & Activity [kBq] & $\gamma$ ray used [keV] & Intensity [\%]  \\
\hline
\hline
     	$^{54}$Mn & 3 & 834.9 & 99.98 \\ 
	$^{137}$Cs & 344 & 661.6 & 85.1 \\
	$^{60}$Co & 183 & 1172 & 99.98 \\ 
 	 &  & 1332 & 99.98 \\ 
	$^{152}$Eu & 310 & 121.8 & 28.53 \\ 
	 &  & 244.7 & 7.55 \\ 
	 &  & 344.3 & 26.59 \\ 
	 &  & 778.9 & 12.93 \\ 

    \hline
        \end{tabular}
\caption{List of radioactive sources used during the efficiency calibration and the $\gamma$ rays used to evaluate the absolute efficiency of the two detectors. All sources were calibrated with an uncertainty of 3\%.}\label{tab:sources}
\end{center}
\end{table}
%%%%%%%%%%%%%%%%%%%%%%%%%%%%%%%%%%%%%%%%%%%%%%%%%%%%%%%%%%%%%%%%%%%%%%%%%
The experimental points have been then fitted with a standard polynomial parametrisation \cite{Xhixha13-JRNC,Marta10-PRC} in order to derive the absolute efficiency  at the energy of the first excited level of $^7$Be, that emits  $\gamma$ rays of 429~keV. In Fig.  \ref{fig:HPGe_efficiency}, the  parametrisation curve for the HPGe is plotted together with the experimental efficiency obtained from the sources. The experimental errors reported in Fig. \ref{fig:HPGe_efficiency} are only statistical and the adopted uncertainty on the fitting procedure is 2\%. The sodium-iodide efficiency is flatter, as shown in Fig. \ref{fig:NaI_efficiency}. We adopted a linear interpolation of the data, obtaining a conservative uncertainty on this method of 3\%. The errors on the parametrisation can be considered independent and should be quadratically combined with the 3\% error on the sources activity.
%+++++++++++++++++++++++++++++++++++++++++++++++++++++++++++++++++++++++++++++++++
%+++++++++++++++++++++++++++++++++++++++++++++++++++++++++++++++++++++++++++++++++
%+++++++++++++++++++++++++++++++++++++++++++++++++++++++++++++++++++++++++++++++++
%+++++++++++++++++++++++++++++++++++++++++++++++++++++++++++++++++++++++++++++++++

The $^{10}$B(p,$\alpha_1 \gamma$)$^{7}$Be yields have been used to obtain the  cross section at the two angles (see Eq. \ref{eq1}). The associated energy in the center of mass has been calculated by using the mean energy definition as in \cite{Caciolli11-AA,Iliadis15-Book}.

%=============================================================================
\section{Discussion}
\label{sec:discussion}
%=============================================================================

The cross section of the $^{10}$B(p,$\alpha_1 \gamma$)$^{7}$Be reaction has been measured in the energy range from 348  to 1795 keV in the center of mass.
The results obtained are presented in Table \ref{tab:cross_section} for the two detectors positions. The data at 90$^\circ$ and 45$^\circ$ are in agreement, supporting the evidence of isotropic angular distribution for the $\gamma$ ray emission at all energies investigated. This statement has been claimed also in previous papers, both at low \cite{Cronin56-PR} and high energies \cite{Chiari16-NIMB-2,Lagoyannis15-NIMB}.

\begin{figure*}[!tbh]
\centering
\includegraphics[width=\textwidth]{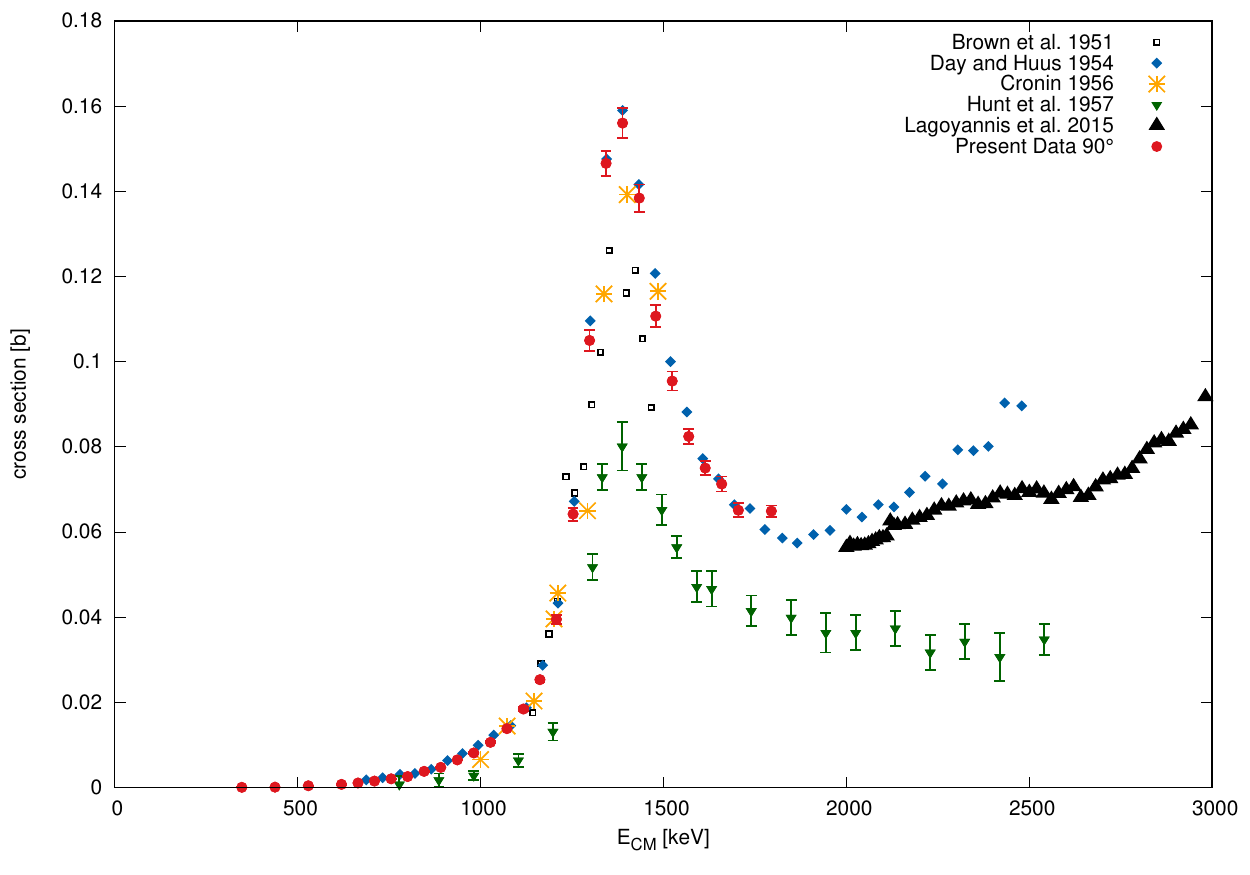}
\caption{The present cross section data shown together with the previous datasets in literature.}
\label{fig:cross_section}
\end{figure*}

In Fig. \ref{fig:cross_section}, the present data (at only one measured angle) are compared with the previous data in literature.
%+++++++++++++++++++++++++++++++++++++++++++++++++++++++++++++++++++++++++++++++++
%+++++++++++++++++++++++++++++++++++++++++++++++++++++++++++++++++++++++++++++++++
%+++++++++++++++++++++++++++++++++++++++++++++++++++++++++++++++++++++++++++++++++
\begin{table}[!h]
\begin{center}
    \begin{tabular}{| c | c | c |}
\hline
$E_{CM}$ [keV]	&	$\sigma(E)$[mb] at 90$^\circ$ 	&	$\sigma(E)$[mb] at 45$^\circ$ 	 	\\
\hline
\hline
348	&	0.05	$\pm$	0.03	&				\\
439	&	0.07	$\pm$	0.05	&				\\
529	&	0.40	$\pm$	0.05	&				\\
620	&	0.72	$\pm$	0.07	&				\\
665	&	1.05	$\pm$	0.08	&				\\
710	&	1.51	$\pm$	0.10	&		\\
756	&	2.03	$\pm$	0.13	&		\\
801	&	2.57	$\pm$	0.19	&		\\
846	&	3.8	$\pm$	0.2&	3.3	$\pm$	0.2	\\
891	&	4.7	$\pm$	0.3	&	4.0	$\pm$	0.3	\\
937	&	6.5	$\pm$	0.4	&	5.7	$\pm$	0.3	\\
982	&	8.1	$\pm$	0.5	&	7.0	$\pm$	0.5	\\
1027	&	10.6	$\pm$	0.7	&	9.1	$\pm$	0.6	\\
1072	&	13.8	$\pm$	0.9	&	12.1	$\pm$	0.8	\\
1117	&	18.4	$\pm$	1.2	&	17.1	$\pm$	1.1	\\
1163	&	25.3	$\pm$	1.5	&	23.1	$\pm$	1.4	\\
1208	&	39	$\pm$	2	&	37	$\pm$	2	\\
1253	&	64	$\pm$	4	&	63	$\pm$	4	\\
1298	&	105	$\pm$	6	&	108	$\pm$	7	\\
1343	&	147	$\pm$	9	&	155	$\pm$	9	\\
1388	&	156	$\pm$	10	&	167	$\pm$	10	\\
1434	&	138	$\pm$	8	&	144	$\pm$	9	\\
1479	&	111 $\pm$	7	&	115	$\pm$	7	\\
1524	&	95	$\pm$	6	&	94	$\pm$	6	\\
1569	&	82	$\pm$	5	&	84	$\pm$	5	\\
1614	&	75	$\pm$	5	&	74	$\pm$	4	\\
1659	&	71	$\pm$	4	&	67	$\pm$	4	\\
1705	&	65	$\pm$	4	&	65	$\pm$	4	\\
1795	&	65	$\pm$	4	&	58	$\pm$	4	\\
\hline
        \end{tabular}
\caption{The cross section values measured in the present experiment at the two angles. The errors reported include both the statistical and the systematic uncertainties.}\label{tab:cross_section}
\end{center}
\end{table}
%+++++++++++++++++++++++++++++++++++++++++++++++++++++++++++++++++++++++++++++++++
%+++++++++++++++++++++++++++++++++++++++++++++++++++++++++++++++++++++++++++++++++
\begin{table}[!h]
\begin{center}
    \begin{tabular}{| l | c |}
\hline
Source	&	Error [\%] 	\\
\hline
\hline
Charge	&	1 \\
Beam energy & 1 \\
Target analysis & 4 \\
radioactive sources & 3 \\
HPGe efficiency & 2 \\
NaI(Tl) efficiency & 3 \\
\hline
Total & 6 \\
\hline
\hline
        \end{tabular}
\caption{Systematic error budget. The component from the parametrisation fit of the efficiency for the two detector is independent and has been summed separately for each of them.}\label{tab:errors}
\end{center}
\end{table}
%+++++++++++++++++++++++++++++++++++++++++++++++++++++++++++++++++++++++++++++++++

The present cross sections are in  agreement with the normalised results reported by Day and Huus  \cite{Day54-PR}.  The present data are also about 20\% higher than Brown et al. \cite{Brown51-PR} and almost a factor of 2 higher than Hunt et al.  \cite{Hunt57-PR}. A clear comparison with the  data from Cronin \cite{Cronin56-PR} is difficult due to the large energy step used in that measurement. That data seem to be lower, but the discrepancy might be due to a slight mismatch in the  energy. At low energies those data are in agreement with the present one. 
%This is actually corroborated by experimental work that compared the experimental thick target yields with what obtained in software simulation using Day and Huus finding an overall good agreement \cite{Mateus07-NIMB}.
The present energy range does not overlap to the data in \cite{Lagoyannis15-NIMB}, therefore, unfortunately, a direct comparison is not possible.

A summary of all sources of systematic uncertainty can be found in Table \ref{tab:errors}.

The $^{10}$B(p,$\alpha_1 \gamma$)$^{7}$Be S-factor from the present experiment has been subtracted from the total S-factor from \cite{Caciolli16-EPJA} to extract the contribution of the $\alpha_0$ channel. The subtraction was done under the assumption of isotropic cross section. A conservative 20\% error was added to the present results to account for the uncertainty on the interpolation of our data.  Results are reported in table \ref{tab:new_Sfactor}.\\
At energies below 1 MeV, the total cross section is dominated by the capture to the ground state, while the capture to the first excited state is negligible. Indeed, the correction to the data from \cite{Caciolli16-EPJA} is only significant at the highest energy measured, where it reaches 11\%. This trend agrees with what expected according to the literature data \cite{Brown51-PR,Cronin56-PR,Rauscher96-PRC}.
%%+++++++++++++++++++++++++++++++++++++++++++++++++++++++++++++++++++++++++++++++++
%\begin{figure*}[!tbh]
%\centering
%\includegraphics[width=0.6\textwidth]{fig6.eps}
%\caption{The comparison of the S-factor for the $\alpha_0$ and $\alpha_1$ channels in the  $^{10}$B(p,$\alpha \gamma$)$^{7}$Be reaction. In the picture the previous works in literature are also presented together with the R-Matrix calculation from \cite{Caciolli16-EPJA}.}
%\label{fig:new_Sfactor}
%\end{figure*}
%%+++++++++++++++++++++++++++++++++++++++++++++++++++++++++++++++++++++++++++++++++
%%%%%%%%%%%%%%%%%%%%%%%%%%%%%%%%%%%%%%%%%%%%%%%%%%%%%%%%%%%%%%%%%%%%%%%%%
\begin{table}[!h]
\begin{center}
    \begin{tabular}{| c | c | c | c |}
    \hline
    $E_{CM}$ & $S$-factor from \cite{Caciolli16-EPJA}  & $S$-factor (only $\alpha_0$) & $\Delta S$   \\
      $[$keV] & [MeV b] & [MeV b] & [MeV b] \\
\hline
\hline
     	249 & 19.96 & 20.0 & 1.2 \\ 
	347 & 14.47 & 14.4 & 0.9 \\ 
	446 & 12.72 & 12.7 & 0.8 \\
	548 & 9.34 & 9.2 & 0.6 \\ 
 	647 & 13.33 & 13.1 & 0.8 \\ 
	749 & 10.75 & 10.4 & 0.6 \\ 
	900 & 16.61 & 15.8 & 1.0 \\ 
	1182 & 20.77 & 18.4 & 1.1 \\ 
	\hline
        \end{tabular}
\caption{Updated S-factor of the $^{10}$B(p,$\alpha_0$)$^{7}$Be reaction compared with what reported in  \cite{Caciolli16-EPJA} The errors are reported only for the updated results.}\label{tab:new_Sfactor}
\end{center}
\end{table}
%%%%%%%%%%%%%%%%%%%%%%%%%%%%%%%%%%%%%%%%%%%%%%%%%%%%%%%%%%%%%%%%%%%%%%%%%
% ===================================================
%=============================================================================
\section{Conclusions}
%=============================================================================

The cross section of the $^{10}$B(p,$\alpha_1 \gamma$)$^{7}$Be reaction has been measured down to center of mass energies of 348 keV, in an energy range typically used for ion beam analysis. Measurements at two different angles provided consistent results, as it was expected according to the literature.\\
The present data agree with the normalized results from Day and Huus \cite{Day54-PR}, that are the most commonly used in ion beam analysis, but the uncertainty has been substantially reduced. Moreover, no scaling procedure needs to be applied to the present data, improving the reliability of the results. \\
Finally, the present results have been used, together with the data from \cite{Caciolli16-EPJA}, to evaluate the relative contribution of the captures to the ground and first excited state.

\subsection*{Acknowledgments}

The authors are indebted to Leonardo La Torre for the help
in setting up the installation and for running the accelerator.
We thank Massimo Loriggiola for the target production
and the LNL mechanical workshops for technical support. We thank Carlo Broggini and Roberto Menegazzo for the fruitful discussions. Financial
support by INFN and University of Padua (Grant
GRIC1317UT) is gratefully acknowledged. 

%
% BibTeX users please use
%\bibliographystyle{epj}
%\bibliography{Caciolli}

\begin{thebibliography}{24}

\bibitem{Mayer-book-1977}
J.~Mayer, E.~Rimini, \emph{Ion Beam Handbook for Material Analysis 1st Edition}
  (Academic Press, 1977)

\bibitem{Nastasi-book-1995}
J.R. Tesmer, M.~Nastasi, \emph{Handbook of Modern Ion Beam Analysis}
  (Pittsburgh, Pa. : Materials Research Society, 1995 and 2nd edition 2010)
 
%\bibitem{Nastasi-book-2010} 
%Yongqiang Wong and Michael Nastasi, \emph{Handbook of Modern Ion Beam Analysis (Volume 2)} 
 % (Pittsburgh, Pa. : Materials Research Society, 2010)

\bibitem{RUMP}
L.R. Doolittle, Nuclear Instruments and Methods in Physics Research Section B:
  Beam Interactions with Materials and Atoms \textbf{9}, 344  (1985)

\bibitem{SIMNRA}
M.~Mayer, AIP Conference Proceedings \textbf{475}, 541 (1999)

\bibitem{Jeynes12-NIMB}
C.~Jeynes, M.~Bailey, N.~Bright, M.~Christopher, G.~Grime, B.~Jones,
  V.~Palitsin, R.~Webb, Nuclear Instruments and Methods in Physics Research
  Section B: Beam Interactions with Materials and Atoms \textbf{271}, 107
  (2012)

\bibitem{Jeynes11-RAST}
C.~Jeynes, R.P. Webb, A.~Lohstroh, Reviews of Accelerator Science and
  Technology \textbf{04}, 41 (2011)
  
\bibitem{IBANDL}
IBANDL database, IAEA (2019), \texttt{http://www-nds.iaea.org/ibandl/}

\bibitem{Sah97-MJ}
R.~Sah, P.~Brown, Microchemical Journal \textbf{56}, 285  (1997)

\bibitem{Mateus07-NIMB}
R.~Mateus, A.~Jesus, M.~Fonseca, H.~Lu{\'\i}s, J.~Ribeiro, Nuclear Instruments
  and Methods in Physics Research Section B: Beam Interactions with Materials
  and Atoms \textbf{264}, 340  (2007)

\bibitem{Moro14-AIPCP}
M.V. Moro, T.F. Silva, G.F. Trindade, N.~Added, M.H. Tabacniks, AIP Conference
  Proceedings \textbf{1625}, 120 (2014)
  
\bibitem{Spitaleri17-PRC}  
C.~Spitaleri, S.~M.~R.~Puglia, M.~La~Cognata, L.~Lamia, S.~Cherubini, A.~Cvetinovic, G.~D'Agata, M.~Gulino, G.~L.~Guardo, I.~Indelicato et~al., Phys. Rev. C \textbf{95}, 035801 (2017)

\bibitem{Caciolli16-EPJA}
{Caciolli, A.}, {Depalo, R.}, {Broggini, C.}, {La Cognata, M.}, {Lamia, L.},
  {Menegazzo, R.}, {Mou, L.}, {Puglia, S. M. R.}, {Rigato, V.}, {Romano, S.}
  et~al., Eur. Phys. J. A \textbf{52}, 136 (2016)

\bibitem{Brown51-PR}
A.~Brown, C.~Snyder, W.~Fowler, C.~Lauritsen, Phys.~Rev. \textbf{82}, 159
  (1951)

\bibitem{Day54-PR}
R.B. Day, T.~Huus, Phys. Rev. \textbf{95}, 1003 (1954)

\bibitem{Cronin56-PR}
J.W. Cronin, Phys. Rev. \textbf{101}, 298 (1956)

\bibitem{Hunt57-PR}
S.E. Hunt, R.A. Pope, W.W. Evans, Phys. Rev. \textbf{106}, 1012 (1957)

\bibitem{Lagoyannis15-NIMB}
A.~Lagoyannis, K.~Preketes-Sigalas, M.~Axiotis, V.~Foteinou, S.~Harissopulos,
  M.~Kokkoris, P.~Misaelides, V.~Paneta, N.~Patronis, Nuclear Instruments and
  Methods in Physics Research Section B: Beam Interactions with Materials and
  Atoms \textbf{342}, 271  (2015)

\bibitem{Iliadis15-Book}
C.~Iliadis, \emph{Nuclear Physics of Stars}, {2$^{\rm nd}$}~edn. (Wiley-VCH,
  Weinheim, 2015)

\bibitem{Caciolli12-EPJA}
A.~{Caciolli}, D.A. {Scott}, A.~{Di Leva}, A.~{Formicola}, M.~{Aliotta},
  M.~{Anders}, A.~{Bellini}, D.~{Bemmerer}, C.~{Broggini}, M.~{Campeggio}
  et~al., Eur.~Phys.~J.~A \textbf{48}, 144 (2012)

\bibitem{Xhixha13-JRNC}
G.~Xhixha, G.P. Bezzon, C.~Broggini, G.P. Buso, A.~Caciolli, I.~Callegari,
  S.~De~Bianchi, G.~Fiorentini, E.~Guastaldi, M.~Ka{\c{c}}eli~Xhixha et~al.,
  Journal of Radioanalytical and Nuclear Chemistry \textbf{295}, 445 (2013)

\bibitem{Marta10-PRC}
M.~{Marta}, E.~{Trompler}, D.~{Bemmerer}, R.~{Beyer}, C.~{Broggini},
  A.~{Caciolli}, M.~{Erhard}, Z.~{F{\"u}l{\"o}p}, E.~{Grosse}, G.~{Gy{\"u}rky}
  et~al., Phys.~Rev.~C \textbf{81}, 055807 (2010)

\bibitem{Caciolli11-AA}
A.~{Caciolli}, C.~{Mazzocchi}, V.~{Capogrosso}, D.~{Bemmerer}, C.~{Broggini},
  P.~{Corvisiero}, H.~{Costantini}, Z.~{Elekes}, A.~{Formicola},
  Z.~{F{\"u}l{\"o}p} et~al., Astron.~Astrophys. \textbf{533}, A66 (2011)

\bibitem{Chiari16-NIMB-2}
M.~Chiari, G.~Ferraccioli, B.~Melon, A.~Nannini, A.~Perego, L.~Salvestrini,
  A.~Lagoyannis, K.~Preketes-Sigalas, Nuclear Instruments and Methods in
  Physics Research Section B: Beam Interactions with Materials and Atoms
  \textbf{366}, 77  (2016)

\bibitem{Rauscher96-PRC}
T.~Rauscher, G.~Raimann, Phys. Rev. C \textbf{53}, 2496 (1996)

\end{thebibliography}
%

\end{document}